\documentclass[twocolumn,floatfix,aps,superscriptaddress]{revtex4}
\usepackage{amsmath,bm}
\usepackage[utf8]{inputenc}
\usepackage[british,UKenglish,USenglish,american]{babel}
\usepackage{bbm}
\usepackage{amssymb}
\usepackage{amsmath}
\usepackage{tikz}\usetikzlibrary{shapes.geometric, arrows}
\tikzstyle{startstop} = [rectangle, rounded corners, minimum width=3cm, minimum height=1cm,text centered, draw=black, fill=red!30]
\tikzstyle{io} = [trapezium, trapezium left angle=70, trapezium right angle=110, minimum width=3cm, minimum height=1cm, text centered, draw=black, fill=blue!30]
\tikzstyle{process} = [rectangle, minimum width=3cm, minimum height=1cm, text centered, text width = 4cm, draw=black, fill=orange!30]
\tikzstyle{decision} = [diamond, minimum width=3cm, minimum height=1cm, text centered, text width = 4cm, draw=black, fill=green!30]
\tikzstyle{arrow} = [thick,->,>=stealth]

\usepackage{graphicx}

\usepackage[normalem]{ulem}
\usepackage{appendix}
\usepackage{color}

\usepackage{pdfpages}
\usepackage{chngpage}

\def\es0{$E_{\rm sym}(\rho_0)$}

\def\us0{$U_{\rm sym}(\rho_0,k_F)$~}

\def\l0{$L(\rho_0)$~}

\newcommand{\beq}{\begin{equation}}
\newcommand{\eeq}{\end{equation}}
\newcommand{\ba}{\begin{array}}
\newcommand{\ea}{\end{array}}
\newcommand{\bea}{\begin{eqnarray}}
\newcommand{\eea}{\end{eqnarray}}
\newcommand{\bi}{\begin{itemize}}  
\newcommand{\ei}{\end{itemize}}
\newcommand{\ben}{\begin{enumerate}} 
\newcommand{\een}{\end{enumerate}}
\newcommand{\bc}{\begin{center}}
\newcommand{\ec}{\end{center}}

\usepackage{pgf}

\usepackage{array}
\begin{document}

\title{Empirical radius formulas for canonical neutron stars from bidirectionally selecting EOS features in extended Bayesian analyses of observational data} 

\author{Jake Richter}
\affiliation{Department of Physics and Astronomy, Texas A$\&$M University-Commerce, Commerce, TX 75429, USA}
\author{Bao-An Li\footnote{Corresponding author: Bao-An.Li@tamuc.edu}}
\affiliation{Department of Physics and Astronomy, Texas A$\&$M University-Commerce, Commerce, TX 75429, USA}
\date{\today}

\begin{abstract}
Given the significant advancement in Bayesian inference of nuclear Equation of State (EOS) from gravitational wave and X-ray observations of neutron stars (NSs), especially since GW170817, is there a data-driven and robust empirical formula for the radius $R_{1.4}$ of canonical NSs in terms of the characteristic EOS parameters (features)? What is the single most important but currently poorly known EOS parameter for determining the $R_{1.4}$? We study these questions by extending the traditional Bayesian analysis which normally ends at presenting the marginalized posterior probability distribution functions (PDFs) of individual EOS parameters and their correlations (or sometimes only the Pearson correlation coefficients which are only reliably useful when the variables are linearly correlated while they are actually often not). Using three regression model-building methodologies: bidirectional step-wise feature selection, Least Absolute Shrinkage Selection Operator (LASSO) regression, and neural network regression on a large set of posterior EOSs and the corresponding $R_{1.4}$ values inferred from earlier comprehensive Bayesian analyses of NS observational data, we systematically and rigorously develop the most probable $R_{1.4}$ formulas with varying statistical accuracy and technical complexity. The most important EOS parameters for determining $R_{1.4}$ are found consistently in each of the feature/model selection processes to be (in order of decreasing importance): curvature $K_{sym}$, slope $L$, skewness $J_{sym}$ of nuclear symmetry energy, skewness $J_{0}$, incompressibility $K_{0}$ of symmetric nuclear matter, and the magnitude $E_{sym} (\rho_0)$ of symmetry energy at the saturation density $\rho_0$ of nuclear matter.

\end{abstract}

\maketitle

\section{Introduction}
Much progress has been made especially since GW170817 in constraining nuclear matter EOS with uncertainty quantification thanks to successful applications of Bayesian statistical tools and various nuclear EOS models in analyzing the new data from both astrophysical observations and terrestrial nuclear experiments \cite{LRP1,LRP2}. In particular, LIGO/VIRGO's observations of gravitational waves from NS mergers and NICER's high-precision observations of X-rays from hot spots on pulsars have helped establish strong correlations between the radii $R_{1.4}$ of canonical NSs of masses around 1.4 M$_{\odot}$ and features of nuclear EOS especially its symmetry energy term, see, e.g. Refs. \cite{Baiotti,BALI19,Rai19,Capano20,Kat20,AngLi,Huth} for recent reviews. 

Within the minimal model of NSs consisting of neutrons, protons, electrons and muons at $\beta$-equilibrium, the core EOS in terms of pressure versus energy density can be obtained from the energy per nucleon $E(\rho,\delta)$ in cold neutron-rich nucleonic matter of density $\rho$ and isospin asymmetry $\delta=(\rho_n-\rho_p)/\rho$ where $\rho_n$ and $\rho_p$ are the densities of neutrons and protons, respectively. The $E(\rho,\delta)$ can be written as \cite{Bom91}
\begin{equation}
E(\rho,\delta)=E_0(\rho)+E_{\rm{sym}}(\rho)\cdot \delta ^{2} +\mathcal{O}(\delta^4)
\end{equation}
where $E_0(\rho)$ is the energy per nucleon in symmetric nuclear matter (SNM) while $E_{\rm{sym}}(\rho)$ is nuclear symmetry energy at density $\rho$.
They can be further parameterized as
\begin{eqnarray}\label{eqn:Es}
E_{0}(\rho) &=& E_{0}(\rho_{0}) + \frac{K_{0}}{2} ( \frac{ \rho - \rho_{0}}{3 \rho_{0}})^{2} + \frac{J_{0}}{6}(\frac{ \rho - \rho_{0}}{3 \rho_{0}})^{3},\\
E_{\rm{sym}}(\rho)&=&E_{\rm{sym}}(\rho_0)+L(\frac{\rho-\rho_0}{3\rho_0})\nonumber\\
&+&\frac{K_{\rm{sym}}}{2}(\frac{\rho-\rho_0}{3\rho_0})^2
  +\frac{J_{\rm{sym}}}{6}(\frac{\rho-\rho_0}{3\rho_0})^3\label{Esym}
\end{eqnarray}
in terms of the incompressibility $K_{0}$ and skewness $J_{0}$ of SNM as well as the magnitude $E_{sym}\equiv E_{sym} (\rho_0)$, curvature $K_{sym}$, slope $L$ and skewness $J_{sym}$ of the symmetry energy $E_{\rm{sym}}(\rho)$ around the saturation density $\rho_0$ of SNM. As we shall discuss in more detail, in building the minimum model for neutron stars, we use the empirical value of $E_{0}(\rho_{0})=-15.9$ MeV at $\rho_0=0.16~{\rm fm}^{-3}$.
These EOS parameters characterize not only properties of neutron-rich matter around $\rho_0$ but also its features away from it. For example, the $E_{\rm{sym}}(\rho)$ and the related pressure in NSs at densities around (2-3)$\rho_0$ that is affecting most strongly the radii of canonical NSs is dominated by the $K_{sym}$ and $J_{sym}$ parameters \cite{BALI19}. 

Have we learned enough from analyzing astrophysical data to establish a direct relationship between the $R_{1.4}$ and the EOS parameters defined above? Such a relationship would be useful for getting a quick check of the astrophysical ramifications of a new value of any of the EOS parameters from new terrestrial experiments or predictions of novel theories without going through the procedure of building a new EOS for NSs and solving again the Tolman-Oppenheimer-Vokolff (TOV) equations. In this regard, it is interesting to note that empirical formulas for the masses and gravitational redshifts of NSs in terms of some of the above EOS parameters have recently been proposed by Sotani {\it et al.} \cite{Sot2,Sot22,Sot3} mostly based on predictions of several tens of nuclear energy density functionals. 

Knowing the relative importance of the above EOS parameters in determining the $R_{1.4}$ has been a longstanding goal of the community. However, the answers to this quest have been strongly model dependent if not controversial \cite{LRP2}. Partially because in the forward-modeling approach by comparing model predictions with observational data, one usually varies one parameter at a time and different parameters generally have very different relative uncertainties. Rarely, any model fully explores its whole multi-parameter space to find the true global minima of the error functions. It is indeed very hard to make a fair comparison of results from varying parameters with very different values and uncertainties to identify globally the most influential EOS parameter on $R_{1.4}$. On the other hand, in the backward Bayesian inference or covariance analyses one often focus on comparing the Pearson correlation coefficients which are only reliably useful when the variables are linearly correlated, in cases where they are actually not linearly related. Thus, an apparently strong correlation between L and $R_{1.4}$ or a large Pearson correlation coefficient between them based on over 500 energy density functional predictions do not necessarily mean the L parameter is most important for determining the $R_{1.4}$. In this regard, a very similar situation exists for the NS crust-core transition density $\rho_t$ which affects significantly the calculation of $R_{1.4}$. As demonstrated and discussed in detail in Refs \cite{BALI19,Zhang2018,NBZ-JPG}, it is actually the $K_{sym}$ that affects most strongly the crust-core transition properties while
very often one simply presents the $\rho_t$ as a function of L as if the latter is the dominating factor determining the $\rho_t$.
In fact, many investigations on the crust-core transition density and pressure, see, e.g., Refs. \cite{JXu2,Xu09,New12,New14,Pro14,Tews17,India17,Holt,Fra-Crust2,Magn}, have shown that once a single experimental or theoretical constraint is applied the EOS parameters will become correlated. In particular, the $K_{sym}$ and L are approximately linearly correlated albeit with model-dependent correlation strength. This correlation can easily mislead people to believe that the L parameter is most important for determining the crust-core transition density. Thus, given the progress made recently in the field and new proposals for future experiments and observations, it is important to answer timely the question on which EOS parameter is most important for determining the $R_{1.4}$.

In this work, using three regression model-building methodologies: bidirectional step-wise feature selection, LASSO regression, and neural network regression on 68,000 posterior EOSs and the corresponding $R_{1.4}$ values inferred from earlier Bayesian analyses of NS observational data in Refs. \cite{Xie19,Xie20a,xi_Li_2021}, we develop a set of  $R_{1.4}$ formulas with varying statistical importance and technical complexity. We found that $K_{sym}$, $L$, $J_{sym}$, $J_{0}$, $K_{0}$ and $E_{sym} (\rho_0)$ are gradually less important for determining $R_{1.4}$.

The rest of the paper is organized as follows. Firstly, for completeness and ease of discussing the main results of this work, we begin by summarizing in Section \ref{s1} the NS EOS model used and the main features of its posteriors PDFs from the earlier Bayesian analyses \cite{Xie19,Xie20a,xi_Li_2021} that are taken as input data in this work. We then discuss in Section \ref{s2} the basics of regression analysis and present an overview of the important statistics. In Section \ref{s3}, three methodologies for regression-model building are discussed: bidirectional step-wise feature selection, LASSO regression, and neural network regression. Here, a suggestion is made to modify the regular bidirectional feature selection process to improve model performance. In Section \ref{s4}, we discuss the results from each of the regression-model building procedures. We demonstrate the advantages of each technique, along with information that is qualitatively different from the traditional correlation analysis. We conclude in Section \ref{s5} with a set of candidate models and a determination of the most important features for predicting $R_{1.4}$. 

\section{Summary of Bayesian posterior EOSs and $R_{1.4}$ values used in this work}\label{s1}
The posterior EOSs and their predictions for $R_{1.4}$ as a list of 
$\{E_{sym}, L, K_{0}, K_{sym}, J_{0}, J _{sym},R_{1.4}\}$ were taken after the Markov Chain Monte Carlo (MCMC) steps have fully reached equilibrium in the Bayesian inference using astrophysical data with known nuclear and astrophysical constraints as priors \cite{Xie19,Xie20a,xi_Li_2021}. The results presented in this work are obtained by using 68,000 posterior EOSs and the corresponding $R_{1.4}$ values. We have checked that this data set is large enough that all of our results are stable.
More specifically, the following radii of canonical NSs were used as independent data: 1) $R_{1.4}=11.9\pm 1.4$ km extracted by the LIGO/VIRGO Collaborations from GW170817 \cite{LIGO18}, 2) $R_{1.4}=10.8^{+2.1}_{-1.6}$ extracted independently also from GW170817 by De {\it et al.} \cite{De18}, 3) $R_{1.4}=11.7^{+1.1}_{-1.1}$ from earlier analysis of quiescent low-mass X-ray binaries observed by Chandra and XMM-Newton observatories \cite{Lattimer14}, and 4)
$R=13.02^{+1.24}_{-1.06}$ km with mass $M=1.44^{+0.15}_{-0.14}$ M$_{\odot}$ \cite{Miller19} or $R=12.71^{+1.83}_{-1.85}$ km with mass $M=1.34\pm0.24$ M$_{\odot}$ \cite{Riley19} for PSR J0030+0451
from NICER Collaboration. The errors quoted are at 90\% confidence level. The likelihood function $P_{\rm radius}$ used is a product of four Gaussian functions, i.e.,
\begin{equation}\label{Likelihood-radius}
 P_{\rm radius}=\prod_{j=1}^{4}\frac{1}{\sqrt{2\pi}\sigma_{\mathrm{obs},j}}\exp[-\frac{(R_{\mathrm{th},j}-R_{\mathrm{obs},j})^{2}}{2\sigma_{\mathrm{obs},j}^{2}}]
 \end{equation}
where $\sigma_{\rm{obs},j}$ represents the $1\sigma$ error bar of the radius from the observation $j$ while $R_{\rm{th},j}$ is the corresponding theoretical prediction. 

The meta-model EOS for NSs consists of a core EOS (described briefly in the introduction) connected smoothly to the NV EOS \cite{Negele73} for the inner crust and the BPS EOS \cite{Baym1971} for the outer crust using the crust-core transition density and pressure evaluated consistently from the core side \cite{Zhang2018} using a thermodynamical approach \cite{Lattimer00,Kubis}. 
As discussed repeatedly in earlier publications \cite{Zhang2018,NBZ-JPG,Xie19,Xie20a,xi_Li_2021,Zhang19,Zhang19b,Zhang20,Zhang20b,Zhang21,Zhang23} involving one of us, once the EOS parameters in Eqs. (\ref{eqn:Es} and \ref{Esym}) are given, a unique EOS for $npe\mu$ matter in neutron stars at $\beta$-equilibrium can be constructed from the energy density of neutron star matter
\begin{equation}\label{lepton-density}
  \varepsilon(\rho, \delta)=\rho [E(\rho,\delta)+M_N]+\varepsilon_l(\rho, \delta),
\end{equation}
where $M_N$ is the average nucleon mass and $\varepsilon_l(\rho, \delta)$ is the lepton energy density calculated from the non-interacting Fermi gas model \cite{Oppenheimer39}. The particle densities (consequently the density profile of isospin asymmetry $\delta(\rho)$) can be obtained by solving the $\beta$-equilibrium condition $\mu_n-\mu_p=\mu_e=\mu_\mu\approx4\delta E_{\rm{sym}}(\rho)$ and the charge neutrality condition $\rho_p=\rho_e+\rho_\mu$. Here the chemical potential for a particle $i$ is calculated from the energy density via
$
  \mu_i=\partial\epsilon(\rho,\delta)/\partial\rho_i.
$
With the above inputs, the barotropic pressure $P(\rho)$ can then be calculated from
\begin{equation}\label{pressure}
  P(\rho)=\rho^2\frac{d\varepsilon(\rho,\delta(\rho))/\rho}{d\rho}.
\end{equation}
Similarly, the  energy density $\varepsilon(\rho, \delta(\rho))$ becomes $\varepsilon(\rho)$ (barotropic) once the density profile of isospin asymmetry $\delta(\rho)$ is obtained. Then the resulting EOS $P(\varepsilon)$ in the form of pressure versus energy density is ready as an input in solving the TOV equations.

It is a meta model (model of models) of the six EOS parameters used for constructing the core EOS and determining the crust-core transition properties. These EOS parameters are generated randomly within their specified prior ranges consistent with our current best knowledge. The resulting EOS models for NSs can mimic essentially all existing EOSs models in the literature. Such meta-model EOS with the minimum assumptions about the compositions of NSs have been widely used in the literature. 
Nevertheless, we notice that such minimum model with only 6 parameters have its limitations. For example, both the saturation density $\rho_0$ and the binding energy of symmetric nuclear matter $E_0(\rho_0)$ (determined mostly by the binding energies and charge radii of finite nuclei) still have some uncertainties although their empirical values are among the only few parameters about which the nuclear physics community has a consensus. The crust structure especially the inner one as well as its connection with the core EOS still have large uncertainties. We adopted the most widely used NV+BPS EOSs for the inner and outer crusts as mentioned earlier in building the minimum model for neutron stars with the smallest number of parameters. As a reference, we notice that a comprehensive Bayesian analyses of atomic masses, charge radii and neutron skins of some finite nuclei together with the recent observations of masses and radii of several neutron stars by LIGO/VIRGO and NICER requires at least 18 parameters in a metal model using the compressible liquid drop model for the crust, a core EOS very similar to ours described above and the extended Thomas-Fermi model for nuclei \cite{Mond23}. It was found necessary to separate the subsaturation and suprasaturation properties of nuclear matter to make the analyses efficient and address some of the tensions between astrophysical observations and terrestrial experiments. The narrowing down of the uncertain prior ranges of $\rho_0$ and $E_0(\rho_0)$ was achieved mostly by using the atomic masses and charge radii alone with little improvement by further incorporating the astrophysical data. 

While it is possible to do a Bayesian analyses with 18 parameters even with the limited data available, to accumulate enough posterior EOSs in the 18 dimensional parameter space in order to do the regression analyses is computationally un-affordable to us presently. Moreover, the sheer number of combinations of 18 parameters to different orders of their multiplications is simply too big for the latest regression approaches to work properly. Thus, among the many still uncertain parameters describing the EOS of neutron stars, we have to choose the most important ones for determining the $R_{1.4}$ (mostly suprasaturation EOS parameters) and make good use of our best knowledge especially about the empirical properties of nuclear matter at $\rho_0$. Given the fact that our minimum model does not consider the possibility of forming baryon resonances (Delta and $N^*$ resonances), hyperons as we all as possible phase transitions, besides the uncertainties of the crustal EOSs and the saturation properties of nuclear matter, we caution the readers that all these uncertainties may have some influences on the $R_{1.4}$. Of course, they all deserve further investigations. Nevertheless, our results presented in this work have their own scientifc merits while they should be understood within the minimum model of neutron stars with the caveat mentioned above.

More details for the particular EOS model used in the Bayesian analysis can be found in Refs. \cite{Xie19,Xie20a,xi_Li_2021}. We note that the peak of the mass-radius sequence for an accepted EOS is required to be at least as high as 1.97 M$_{\odot}$ (the same condition that LIGO/VIRGO used in analyzing GW170817). Moreover, all accepted EOSs are dynamically stable throughout the entire NS and always remains casual. This particular meta-model EOS has been used extensively in studying many internal properties and observables of NSs, see, e.g., Refs. \cite{Zhang2018,Zhang19,Zhang19b,Zhang20,Zhang20b,Zhang21}. 

\begin{table}[h]
\caption{Mean values and standard deviations of the six EOS parameters and the corresponding $R_{1.4}$ values obtained using their posterior distributions constrained by neutron star observations in a Bayesian analysis done in Ref. \cite{xi_Li_2021}.}
\vspace{0.2cm}
\label{tb:eosStats}
\begin{tabular}{c|c|c}
\hline
 EOS Parameter ($x_i$) & Mean ($\Bar{x_i}$)& Standard Deviation ($\sigma_i$)\\\hline
$K_{sym}$ & -158.97 & 80.45 \\  
$L$ & 54.13 & 13.63 \\ 
$J_{sym}$ & 473.01& 247.72 \\  
$K_{0}$ & 239.95 & 11.52 \\ 
$ J_{0}$ & -163.01 & 69.07 \\ 
$E_{sym}$ & 31.87 & 1.84 \\ 
$R_{1.4}$& 12.10 & 0.38 \\ 
\hline
\end{tabular}
\end{table}
The PDFs of the six EOS parameters and their correlations can be found in Ref. \cite{xi_Li_2021}. As one of the direct inputs for the analyses in this work, shown in Table \ref{tb:eosStats} are the posterior mean $\Bar{x_i}$ and standard deviation $\sigma_i$ for each of the six EOS parameters $x_i$ with $i=1~{\rm to}~6$. We note the radius is constrained to around 12 km and $E_{sym}$ having the lowest standard deviation is the most constrained EOS parameter. $K_{0}$ and $L$ are the next most constrained parameters which demonstrate current strong beliefs about their values. The remaining parameters $J_{0}$, $K_{sym}$ and $J_{sym}$ have large variations. In fact, the uncertainties are large enough that for some of these parameters it is not surely known whether they carry a positive or negative sign. We emphasize that both the means and standard deviations of the symmetry energy parameters are all in excellent agreement with the current world averages based on over 80 independent analyses of various nuclear and astrophysical data including 24 Bayesian analyses by different groups of essentially the same astrophysical data available since GW170817 \cite{LIBA21,NBZ23}. The statistics of $E_{sym}$ and $L$ are also in very good agreement with predictions of the latest chiral effective field theory \cite{eft}. The latter, however, currently cannot predict accurately the high-density parameters $J_{0}$, $K_{sym}$ and $J_{sym}$. 

For the present work seeking the most probable formulas for $R_{1.4}$, the results listed in Table \ref{tb:eosStats} will be used as the basis to construct the reduced variable 
\begin{equation}\label{redu}
x_i (\rm{reduced})\equiv (x_i-\Bar{x_i})/{\sigma_i}
\end{equation}
with $i=1~{\rm to}~ 6$ to treat all EOS parameters on equal footing, and this is also necessary for several other considerations as we shall discuss in the next Section. Namely, all EOS parameters and $R_{1.4}$ are made dimensionless by first centering them to have a mean value of 0 and then scaling them to have a variance of 1. We then take the six reduced EOS parameters as the basic EOS features. In addition, different combinations of them will then be constructed as candidates to expand the total number of features to be used in the regressions models. 
 
 \begin{figure}
 \vspace{-1cm}
    \includegraphics{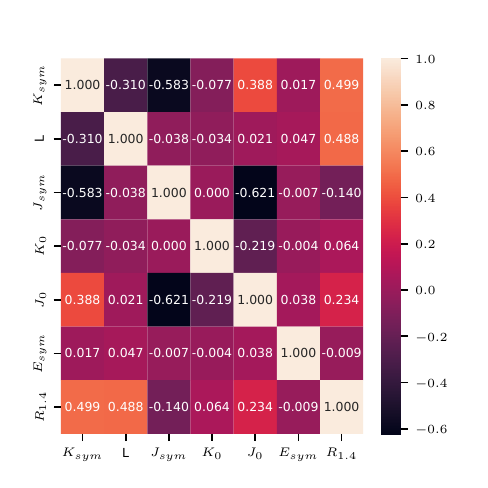}
    \caption{ Pearson Correlation Coefficients (numbers in each box) between the original EOS parameters and $R_{1.4}$. Darker purple (lighter orange) colors indicate high negative (positive) correlations.}
    \label{fig:corrHeat}
\end{figure}
Shown in Fig. \ref{fig:corrHeat} are the corresponding Pearson correlation coefficients for $\{E_{sym}, L, K_{0}, K_{sym}, J_{0}, J _{sym},R_{1.4}\}$ we obtained using data from the same Bayesian analysis \cite{xi_Li_2021}. It is seen that $K_{sym}$ has the highest correlation with $R_{1.4}$ and is expected to be the most informative parameter. However, by its correlation score it is only slightly more correlated with $R_{1.4}$ than $L$ with a difference in scores less than $0.01$. We also note that $K_{sym}$ and $L$ are negatively correlated with each other. $J_{0}$ and $J_{sym}$ are the next most correlated with $R_{1.4}$ and both exhibit higher correlations with $K_{sym}$ than they do with $R_{1.4}$. Finally, $K_{0}$ and $E_{sym}$ are the most uncorrelated with $R_{1.4}$, having near zero-correlations with $R_{1.4}$ and many of the other parameters. The lack of correlation between $K_{0}$, $E_{sym}$ and $R_{1.4}$ is likely because they are the most constrained parameters and imposition of a stricter prior distribution leads to smaller variations in their values. Correlations between the remaining parameters and $R_{1.4}$ are more intractable since they all vary significantly during the Bayesian inference. 

We emphasize that the Pearson correlation coefficients shown in Fig. \ref{fig:corrHeat} cannot encapsulate non-linear relationships between variables. These correlation coefficients may undervalue the strength and nature of the relationships and some dependencies of the variables may be misleading if their underlying correlations are nonlinear. We can remedy these issues by performing relatively simple regression analyses and use feature selection techniques. In the process, we can use both parametric correlation methods and non-parametric (e.g., neutral network) ones. The latter often need less assumptions about the distributions of the variables but have no precise parametric formulas accompanying the strength of the relationship \cite{verma}. 

\section{ Regression Preliminaries}\label{s2}
Generally, a linear regression model \cite{neter_regression,murphy_machine} asserts a functional relationship
\begin{equation}
\hat{r}_{k}  = \sum_{i}  \beta_{i} x_{i,k} + \epsilon_{k} = \vec{\beta}\vec{X}_{k} + \epsilon_{k}
\label{eqn:regModel}
\end{equation} 
between an observable $\hat{r}_{k}$ and a set of features $\{\vec{x_{i}}\}_k$ where $k$ is the sample index. In the present study, $\hat{r}_{k}$ is the regression-model predicted value of $R_{1.4}$ using the $k^{th}$ EOS parameter set (features) $\{\vec{x_{i}}\}_k=\{E_{sym}, L, K_{0}, K_{sym}, J_{0}, J _{sym},R_{1.4}\}_k$ with $k$ running from 1 to 68,000 through the entire EOS ensemble. $\beta_{i}$ are the coefficients for the $i^{th}$ feature found by minimizing the total distance between observations (actualization) and regression-model predictions
\begin{equation}
D_{ls}  = \sum_{k}( r_{k} - \hat{r}_{k} )^{2}
\label{eq:leastSquares}
\end{equation}
where $r_{k}$ is the TOV predicted $R_{1.4}$ value (defined as an ``observation'' in the context of this study) with the $k^{th}$ EOS parameter set from the Bayesian analysis. The $\beta_{i}$ are considered as the parameters of the regression model. We note here that as we are going to consider combinations of the original six EOS parameters, the $i$ here is general and can be much larger than 6. 

The difference between the observation (actualization) $r_{k}$ and the regression-model prediction $\hat{r}_{k}$ is a residual and is considered an error $ \epsilon_{k} $ of the regression-model prediction for the $k^{th}$ sample. It is important that the errors be normally distributed and have a constant variance for all predictions \cite{neter_regression}, i.e. homoscedastic. Although, linear regression is robust to violations of these assumptions when there are many observations, or the violations in these assumptions are relatively minor. To ensure violations are moderate and improve model generalizability modifications can be made to linear regression models, as with the Least Absolute Shrinkage and Selection Operator (LASSO) regression \cite{buhl_lasso} which minimizes
\begin{equation} 
    D_{lasso} = \sum_{k}( r_{k} - \hat{r}_{k})^{2} + \lambda \sum_{i}^{n} |\beta_{i}| .
    \label{eq:lassReg}
\end{equation}
Adding the absolute values of all the regression-model parameters $\beta_{j}$ penalizes 
large coefficients weighted by a learning parameter $\lambda$. Here $\lambda$ is a real valued number in the interval [0,1] and affects how quickly $\beta_{i}$ coefficients drop off in comparison with the $\beta_{i}$ found in least squares regression. Note that $\lambda = 0$ is equivalent to minimizing $D_{ls}$ and a $\lambda = 1$ forces all the coefficients to be 0. In cases when $\lambda \neq 1$ minimizing $D_{lasso}$ can still cause some, or all, $\beta_{i}$ coefficients to become 0 and thus nullify the effects of the features they lead. When some $\beta_{i}$ do become 0 then LASSO regression can be used as feature selection method by keeping only the features with non-zero coefficients. More generally, LASSO regression can prevent models from overifitting data and can increase model generalizability. 

To ease the following discussions, we recall in the following a few terminologies necessary. Two metrics are commonly used to determine the quality of a regression model of the form in Eq. (\ref{eqn:regModel}).  
The total sum of squares (TSS) 
\begin{equation}\label{eqn:TSS}
\text{ TSS } \equiv \sum_{k} ( r_{k} - \bar{r} )^{2}.
\end{equation}
is the total distance between all the actualized observable $r_{k}$ 
from their mean $\bar{r}$ (completely determined by the Bayesian analysis in this study). While the sum of squared errors (SSE) 
\begin{equation}\label{eqn:SSE}
\text{ SSE } \equiv \sum_{k}(r_{k} - \hat{r}_{k})^{2}=D_{ls}.
\end{equation} 
is the total distance between all the predictions of a regression model and the corresponding observations.


For model selections, we use the following three statistical measures of the quality of a regression model:
\begin{itemize}
    \item 
The coefficient of determination \cite{wackerly_stats, neter_regression} ($R^{2}$) \begin{equation}\label{eqn:r2}
R^{2} \equiv 1 - \frac{\text{SSE}}{\text{TSS}}
\end{equation}
measures the explained variance of a model. For a given data set and regression-model, SSE can be 0.0, indicating a perfect fit. However, a model can also be arbitrarily poor leading to an SSE greater than TSS and achieve an arbitrarily lower and negative $R^{2}$ values. Thus, $R^{2}$ has a maximum value of 1 but is unbounded below. 

\item
The Akaike Information Criterion \cite{akaike_aic} (AIC)
\begin{equation}
\text{ AIC } \equiv   q - \text{Ln}(\ell ) = 2 q - n\text{Ln}(\frac{\text{SSE}}{d})
\label{eqn:AIC}
\end{equation}
is a model selection statistic which takes account for the number of parameters a model contains and the model's fit as measured by the maximum likelihood estimate. The first expression in Eq.\ (\ref{eqn:AIC}) is general for any 
maximum likelihood estimate $\ell$, and the final one is the expression for AIC when $\ell$ is determined by the least squares minimization. q is the number of estimated parameters in the model, n is the number of observations, and $ d $ is the degree of freedom in the model. While the coefficient of determination $R^{2}$ provides useful information regardless of the existence of other models, 
the AIC can only be used to compare the quality between models that are fit to the same data set. In a given set of models, the model with the smallest AIC is considered the best one. 
\item
The partial F-statistic \cite{neter_regression} ( $F^{*}$ )
\begin{equation}\label{eqn:partialF}
F^{*} \equiv \frac{(SSE_{\nu} - SSE_{f}) d_{f}}{SSE_{f}( d_{\nu} - d_{f})} 
\end{equation} compares the quality of a full model with a nested version of itself.
A full model is defined by the modeller to include as many features of interest and a nested model must contain less features than the full model, 
all of which must be in the full model, i.e. the set of features in the nested model should be a proper subset of the features in the full model. In Eq.\ (\ref{eqn:partialF}),  $SSE_{f(\nu)}$  and 
$d_{f(\nu)}$ are the sum of errors squared and degrees of freedom for the full (nested) model, respectively.

The $F^{*}$ statistic follows a one tailed F distribution and can be written as a ratio of two $\chi^{2}$ distributions \cite{casella_stats,neter_regression}. It is used to test 
the null hypothesis that the set of parameters $ \{ \beta_{j} \}$ in a full model, which are not in the nested model, are simultaneously 0. A higher 
F-statistic yields a lower probability (p-value) in the F-distribution and is evidence that at least one of $\beta_{k}$ 
in the set $\{ \beta_{j} \} $ being tested is non-zero, i.e. evidence to reject the null hypothesis. Acceptable p-values, i.e. significance levels are set before modelling and if a p-value is greater than the significance level, then $F^{*}$ is not considered to have provided sufficient evidence against the null hypothesis and the set $\{ \beta_{j} \} $ would be assumed simultaneously 0. If the full model includes only one extra parameter 
than the nested model, then the null hypothesis can be used to test whether a single $\beta_{i} = 0$, thus providing evidence for the statistical significance of a single parameter. 
\end{itemize}

\begin{figure*}
\centering
\includegraphics[angle = 0.0, scale = 1.0]{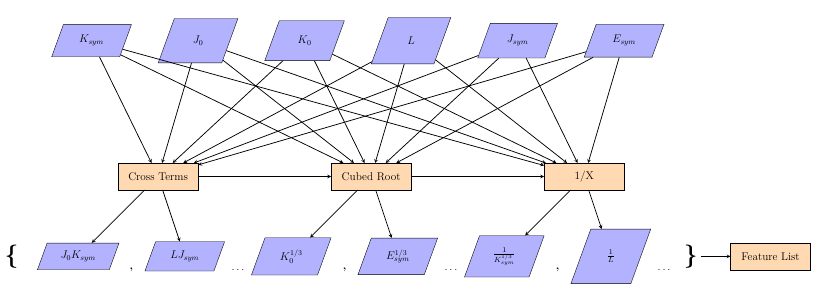}
\caption{Diagrammatic representation of the feature generating procedure before correlation filtering is applied. }
\label{fig:featGen}
\end{figure*}
\section{ Methods }\label{s3}
First of all, we note that unless otherwise specified, from this point forward we discuss the relationship between EOS dimensionless features and the dimensionless radius $R_{1.4}$. All the symbols in the plots and tables are reduced variables as defined in Eq.\ (\ref{redu}).

As outlined in Fig.\ \ref{fig:featGen}, 
each EOS parameter, their polynomial terms up to degree six, along with their cubed root, reciprocals, and logarithms are considered as features of interest. Most of the features are  
generated to take account possible complexity of different terms, therefore to avoid an overly complex model each feature is checked for correlation with every other feature. If the correlation between two features is greater 
than 0.85 as measured by the Pearson correlation coefficient the feature which has higher complexity is removed from the list of generated features. Complexity is determined by the abstraction from the original EOS parameters, and thus the latter are defined as features with the lowest order complexity. Logarithms have the highest complexity due to the required positive argument, fractions have the second highest complexity due to their divergence at 0, cubed root terms are then third highest complexity and finally the cross terms are considered the least complex generated features.

 Three different analyses are performed after feature generation and filtration: Bidirectional feature selection, LASSO regression, and neural network regression. We use the Python package \textit{sci-kit learn} \cite{scikit} to do least squares regression, LASSO regression, and for building the architecture of the neural network. 
 
\begin{figure*}
\centering
\includegraphics[angle=-90, scale = .5]{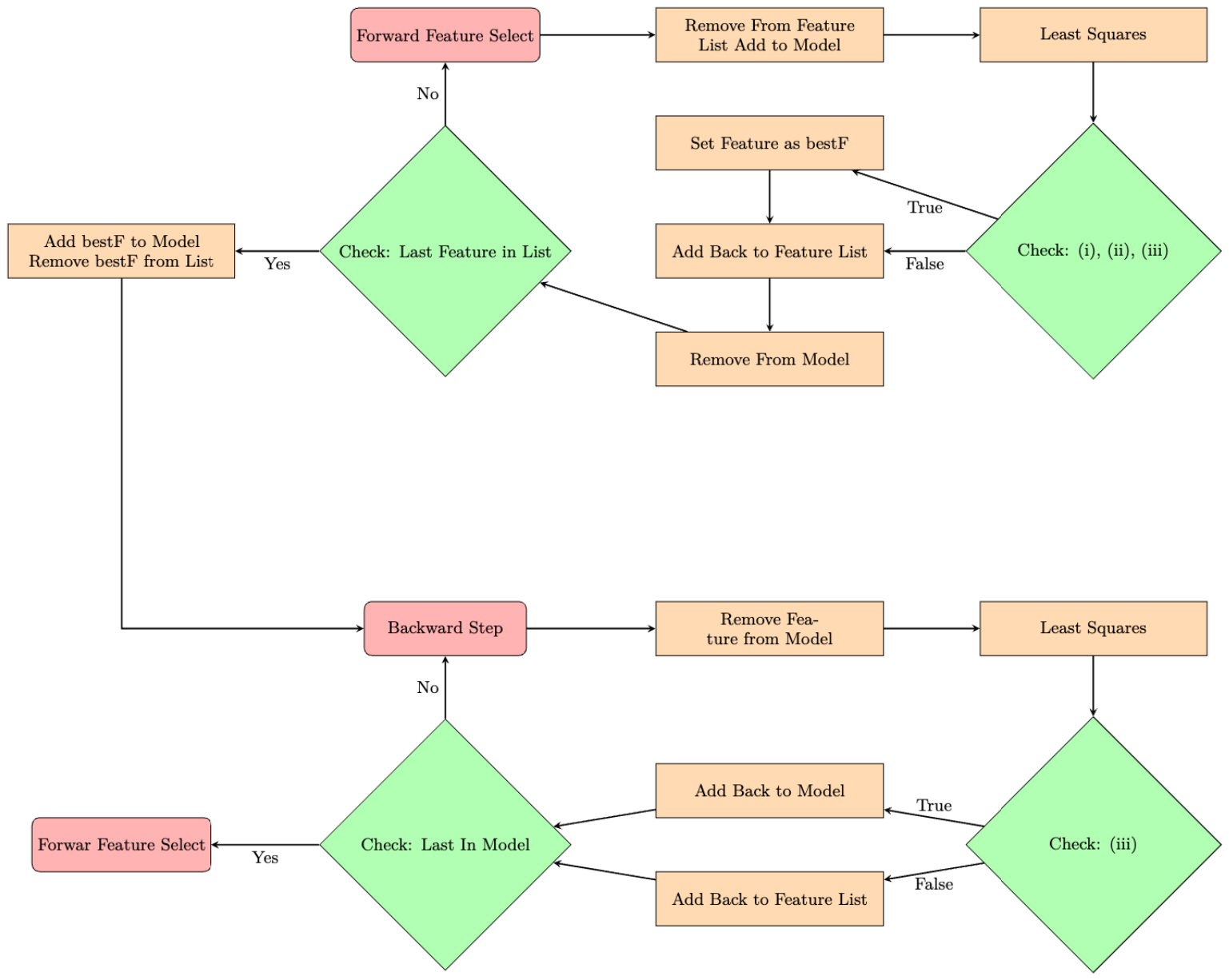}
\caption{Diagrammatic representation of the Modified Bidirectional Feature Selection Algorithm. The numerals (i), (ii), and (iii) are the conditions discussed in the methods section. Removing condition (ii) would give the regular bidirectional selection procedure. Beginning with a forward step a model is built by one-at-a-time selection and fitting to find the best feature at the given model step. After each forward step is taken a 'backward step' is performed to ensure each feature in the model is statistically significant, if not it will be removed. The process terminates when no 'bestF' is found in the forward step process.}
\label{fig:forwardDiagram}
\end{figure*}

\subsection{Parametric Regression-Model Building}
A bidirectional step algorithm as illustrated in Figure \ref{fig:forwardDiagram}, similar to the one described in Ref. \cite{neter_regression} is implemented to decide which of the features will create 
an efficient model. First, a working model containing only the average value of the dimensionless radius distribution 
is considered and because it has been centered to 0.0, this is equivalent to a completely 
empty model. Then a ``forward step" is taken, each feature in the generated set 
is considered for addition into the model by temporarily adding the feature, performing least squares minimization, and then removing it from the model. 
The parameter that (i) has the highest partial F-statistic (ii) improves or 
maintains model residual statistics as measured by the residual distribution skew and kurtosis (third and fourth statistical moments, respectively), and (iii) has an $F^{*}$ p-value greater than 0.05 will be considered significant, added to the working model and removed from the feature list. 
After adding a feature to the model a ``backward step" is taken wherein each feature in the current working model is removed one at a time to determine if they are statistically significant by the $F^{*}$ p-value. If it is determined that there are insignificant features in the backward step then they are removed from the model. After the backward step, the procedure returns to perform a forward step and iteratively builds a model. This process of forward and backward stepping repeats until no other features
can be added to the model either because all have been included or there is not another parameter which satisfies conditions (i),(ii) and (iii) in the forward step.  

A significance level of 0.05 is chosen here to  ensure that realistic variables are included in the model. Higher significance levels may lead to less significant features being included and create unnecessarily complex models. Lower significance levels
may exclude informative variables that would improve model performance. It should 
be noted that the condition (ii) is not commonly used in the literature. To our best knowledge, a condition similar to it has not been imposed in a step wise algorithm such as this one. The goal of condition (ii) is to force selecting features that keep the residuals distributed approximately symmetric about the prediction. Thus, when using the condition (ii) we consider this to be a modified bidirectional step algorithm. 

The LASSO penalized regression is performed on the same set of features described in the above generation process. The learning rate $\lambda$ is set as a hyper parameter of the model selection process and chosen so that AIC is minimized. In the Bidirectional feature selection and LASSO regression, we consider different sets of generated features. The feature sets are characterized by the highest polynomial terms that were generated before filtering out features with excessive correlation, e.g. a feature set with only the original EOS parameters (and their expressions) as generated features is considered the complexity degree one feature set. This allows us to systematically study the effects of including higher complexity terms by varying the complexity of terms in the generated feature set. This is  especially true when analyzing the bidirectional selection techniques which provide step-by-step model selections that are more sensitive to different correlations. In the LASSO regression, changing complexity of the feature set may only change the number of parameters with non-zero coefficients.

\subsection{Neural Network (non-parametric) Regression}
Neural networks provide a method for performing regression analysis with arbitrarily high degrees of complexity and low computational cost \cite{murphy_machine}. Networks are composed of layers and each layer is composed a set of neurons. The first layer is the input layer which is followed by a sequence of hidden layers and finally an output layer which gives the networks predictions. Each neuron is a function (usually a non-linear function) that takes as an input a weighted sum of the outputs from the layer preceding it and gives a single valued output for the next layer. Weights between layers are then adjusted to improve model performance in the output layer.  In this work a neural network is trained on the Bayesian posterior data set discussed earlier. We use the sci-kit learn's MLPRegressor class to build a simple multilayer perceptron for the regression problem. The network consists of the six EOS parameters as an input layer and two hidden layers with six neurons each. All hidden layer neurons were activated by the hyperbolic tangent function. Training was done to minimize the mean squared error (MSE)
\begin{equation}
\text{MSE} \equiv \frac{1}{n}\sum_{k}^{n}( r_{k} - \hat{r}_{k} )^{2}
\label{eq:mse}
\end{equation}
where n is the number of observations. The data is split 75/25 so that 75$\%$ of it is used for training, the rest is for model testing. For updating the weights, we use the \textit{adam} method with a batch size of 200 samples with 1000 epochs. 

After training is complete we use the Python package \textit{eli5} to perform permutation importance testing on the input layer features. The procedure chooses one feature vector to permute and passes it through the neural network with the other features remaining unaltered. The permutation of a single feature will alter the predictions and change the MSE. The difference in MSE between the original network with no permuted features and the network with a permuted feature is calculated and stored. The permuted feature is then returned to its original state and another feature is chosen to be permuted, again, the difference in MSE as a result of the permutation is calculated and stored. When all the features have been permuted once we compare the change in MSE to determine which feature the neural network was most reliant on. A higher change in MSE indicates the neural network was more dependent on the given feature since its permutation lead to a more significant loss in information. Thus, a feature with higher change in MSE after permutation would be considered a more important predictor.

\section{ Results and discussions}\label{s4}
Here we present results from applying the regression and feature selection techniques described above. Many candidate models are found and we discuss their efficacy via model selection statistics and their residuals. Finally, a series of suggested $R_{1.4}$ models are presented which best characterize the data set from the Bayesian analysis.  

\subsection{Bidirectional EOS Feature Selections}
\begin{figure*} 
    \includegraphics[scale = .85]{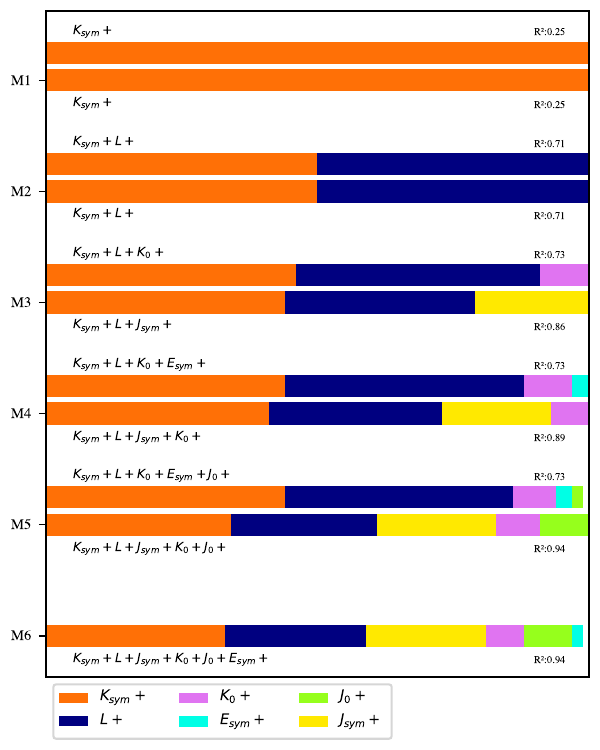}
    \caption{Models selected by the bidirectional feature selection processes when degree 1 terms and expressions are generated. Each $Mi$ (i = 0,1,2...) is a step in the feature selection process  from either the modified procedure or the regular procedure. Top bars are the models selected when condition (ii) is imposed and the bottom bars are the selected models when condition (ii) is not imposed. Colors of the bar are related to a feature that was used in the models, and their lengths are proportional to the its coefficient size as a fraction of the sum of all the coefficient sizes. }
    \label{fig:parModel}
\end{figure*}

Shown in Fig.\ \ref{fig:parModel} are the results of the bidirectional feature selections made at each step where the only features being considered are the original EOS parameters, their cubed roots, logarithms, and reciprocals, i.e. no cross terms. For clarity and brevity, when discussing the models with $i$ features selected from a feature set with generated polynomial terms up to degree $j$ in the modified (regular) process we denote this as M$i^{j}_{mod (reg)}$. The top bar of each M$i$ are the features selected by the  modified bidirectional selection procedure (when  condition (ii) is imposed) and bottom bars are the features selected by the regular bidirectional selection process. The length of each color in the bar is proportional to $
|\beta_{i}|/\sum_{i}|\beta_{i}| 
$ where $\beta_{i}$ coefficient leads the indicated parameter.

The first steps shown in Fig.\ \ref{fig:parModel}, $M1^{1}_{reg}$ and $M1^{1}_{mod}$ choose $K_{sym}$  due to it having a higher Pearson correlation coefficient with $R_{1.4}$ than all other EOS parameters as shown in Fig.\ \ref{fig:corrHeat}. Subsequently, $ L $ is added to both models. The resulting coefficient for L is slightly less than that for $K_{sym}$.  The $R^{2}$ score
increases from $ 0.25 $ in $M1^{1}_{mod}$ to $0.71$ in $M2^{1}_{mod}$. In the modified process the next features added (in order) are $K_{0}$, $J_{0}$ and $E_{sym}$. After the addition of $K_{0}$, there is no significant increase in $R^{2}$ and the coefficients leading $J_{0}$ and $E_{sym}$ are small in comparison to 
$K_{0}$ and especially when compared with $K_{sym}$ and $L$. 

In the regular bidirectional selection process, $M3^{1}_{reg}$ differs from $M3^{1}_{mod}$ in the choice of $J_{sym}$. We note this choice significantly improves the $R^{2}$ score of $M3^{1}_{reg}$ over $M3^{1}_{mod}$, with a difference in scores of $~0.13$ in their third steps. The last three parameters added in the regular bidirectional selection
process are: $K_{0}$, $J_{0}$ and $E_{sym}$. They are the same as the last three parameters added in the modified process. The final model contains all the original EOS parameters with $R^{2}=0.94$. Both selection methods terminate automatically without including any logarithms, reciprocal terms, or cubed roots. The logarithms and cubed roots were considered but then discarded in the feature building process due to their high correlations with the original EOS parameters. The reciprocal terms of $K_{sym}$, $J_{sym}$, $L$ were all uncorrelated with the original parameters and were eligible to be included in either step wise selection algorithm. Their exclusion in both step wise selection processes indicates there is no great explanatory power of the reciprocal terms and they provide no improvement in the distribution of the residuals.  

Moreover, the regular bidirectional selection process finds a more accurate model as measured by $R^{2}$ where the final models differ in $R^{2}$ by $0.26$. The difference in accuracy can be attributed to the exclusion of $J_{sym}$ in the modified procedure. When $J_{sym}$ is included in the regular feature selection process, $K_{0}$ and $J_{0}$ provide more significant increase in $R^{2}$ when compared to their effects on $R^{2}$ in the modified procedure which does not include $J_{sym}$. This suggests $J_{sym}$ has a confounding effect on $K_{0}$ and $J_{0}$ and should be included in an appropriate model of $R_{1.4}$. However, whether $J_{sym}$ should be included as a stand-alone linear term, or in higher polynomial terms is unclear. It is interesting to note here that Bayesian analyses 
\cite{Xie19,Xie20a,xi_Li_2021} have clearly indicated that the currently available NS observational data do not provide a tight constrain on $J_{sym}$. With or without it in parameterizing the EOS for Bayesian analyses affects the PDFs of other EOS parameters similar to the above findings.

\begin{figure*}
\centering
\includegraphics[scale = .85]{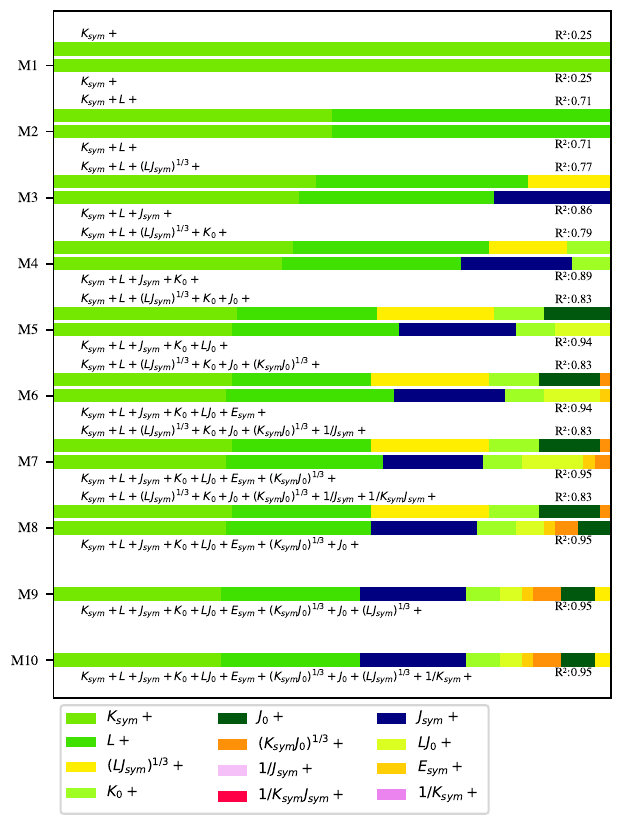}
\caption{Models selected by bidirectional feature selection when degree two cross terms and expressions are considered. Steps, bar lengths and bar colors are determined as in Fig. \ref{fig:parModel} with different color palette being applied. }\label{fig:parModel2}
\end{figure*}

Shown in Fig.\ \ref{fig:parModel2} are the dynamics of the bidirectional step-wise selection when degree two polynomial terms 
and their expressions are generated. Again, $K_{sym}$ and $L$ are the first two parameters added in the modified and unmodified procedure. In the modified procedure, as more parameters are added 
the fractional weights of $K_{sym}$ and $L$ consistently  decrease until the fifth parameter is added. The first large fractional decrease in $L$ is when $M3^{2}_{mod}$ includes $(LJ_{sym})^{\frac{1}{3}}$, also at $M3^{2}_{mod}$, there is a slight decrease in the weight of $K_{sym}$. Since $(LJ_{sym})^{\frac{1}{3}}$ is almost directly correlated with $L$ and $J_{sym}$ is non trivially correlated with $K_{sym}$, the trade-off in coefficient size is expected. We note that $M3^{2}_{mod}$ has a higher $R^{2}$ score than $M3^{1}_{mod}$. It indicates that non-linear expressions of explanatory features which have already been included in the model, may provide more information than including original EOS features like $K_{0}$ and $J_{0}$. The latter two are the next parameters included in the model, and the increase in $R^{2}$ score is greater than the increase seen when only first degree terms were included. This adds to the prior evidence that $J_{sym}$ has positive confounding effect on the variables $K_{0}$ and $J_{0}$ since $J_{sym}$ appears in a cross term with $L$ and is likely the reason $R^{2}$ increases.

In the unmodified case, step wise selection finds models up to  $M4^{2}_{reg}$ are the same as models up to $M4^{1}_{reg}$. $M5_{reg}^{2}$ selects $L J_{0}$ as the first cross term in the regular bidirectional procedure. After $LJ_{0}$ there is no significant increase in $R^{2}$. When $(K_{sym} J_{0})^{\frac{1}{3}}$ are included in $M7^{2}_{reg}$, the $R^{2}$ reaches near its maximum value and the fractional coefficients are significantly smaller than the original EOS features. Notably, the fractional weights of $K_{sym}$, $L$, and $J_{sym}$ are mostly constant with the addition of extra features and most of the trade-off between feature coefficient sizes occur between higher complexity terms.

\begin{figure*}
\includegraphics[scale =1.0]{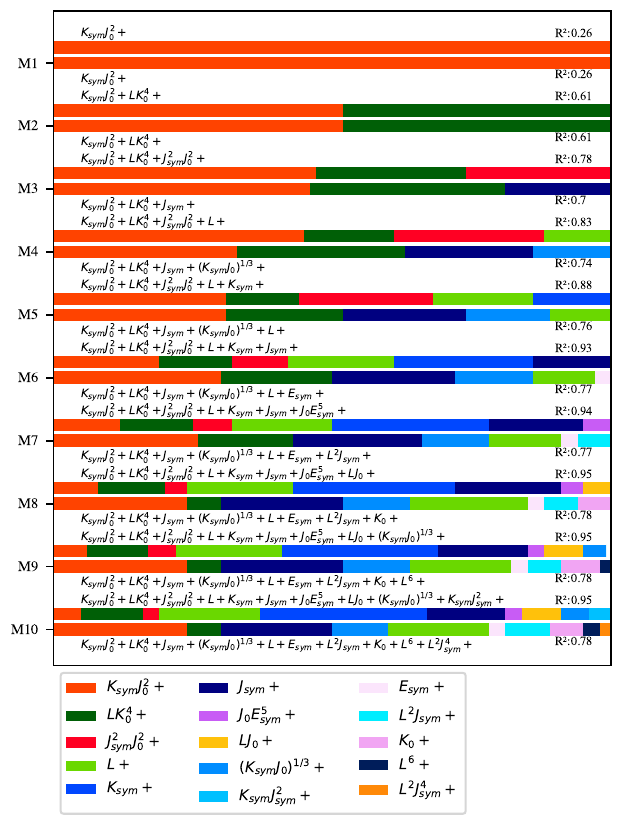}
\caption{Models selected by the bidirectional feature selection when degree six cross terms and expressions are considered. Steps, bar lengths and bar colors are determined as in Fig. \ref{fig:parModel} and Fig. \ref{fig:parModel2} with different color palette being applied. Here, each top bar at an $M{i}$ is the unmodified selection process and the bottom bar is the selection from the modified procedure. }
\label{fig:deg6Dynam}
\end{figure*}

In considering only the lower complexity degree terms, the bidirectional feature selection algorithm shows the progression of the relationships between the original EOS parameters and $R_{1.4}$. 
Here, $K_{sym}$ and $L$ are found to be the most important predictors. Moreover, a comparison of their fractional coefficients indicates that $K_{sym}$ can be considered the most important parameter for determining $R_{1.4}$ in models which only include low complexity terms, regardless of whether a modified step wise algorithm is considered. Additional information from our analyses indicates that $J_{sym}$ is also important for predicting radius. In fact, $J_{sym}$ is more important than its correlations would suggest. In Fig.\ \ref{fig:corrHeat} $J_{sym}$ has a lower correlation with $R_{1.4}$ than $J_{0}$, but it tends to increase the accuracy in modeling $R_{1.4}$ more than $J_{0}$. 
This finding is actually consistent with our earlier knowledge \cite{BALI19}. The $J_{sym}$ controls mostly the high-density symmetry energy while the $J_{0}$ controls mostly the symmetric nuclear matter EOS at densities above about $2\rho_0$ \cite{BALI19}. They are known to mainly affect separately the radii and masses of NSs. This physics is naturally in the Bayesian posteriors. It is reassuring that the feature selection techniques revealed it numerically and automatically. 

The three EOS parameters $K_{0}$, $E_{sym}$, $J_{0}$ are nearly unimportant in predicting $R_{1.4}$
and only provide small contributions in improving accuracy when $J_{sym}$ is included. The unmodified bidirectional selection process indicates that $J_{sym}$ is an important predictor for an accurate model. However, its exclusion in the modified procedure indicates there need to be correcting features in the model to offset its effect on the distribution of the residuals. Again, this is due to the fact that available NS observational data do not constrain the high-density symmetry energy parameter $J_{sym}$. Its value fluctuates irregularly in the Bayesian posteriors \cite{Xie19,Xie20a}.

\begin{figure*}[ht]
\includegraphics[scale = .55]{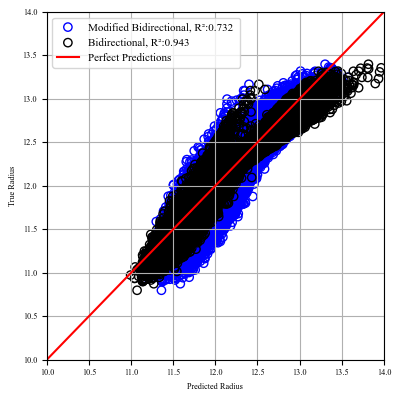}
\includegraphics[scale = .55]{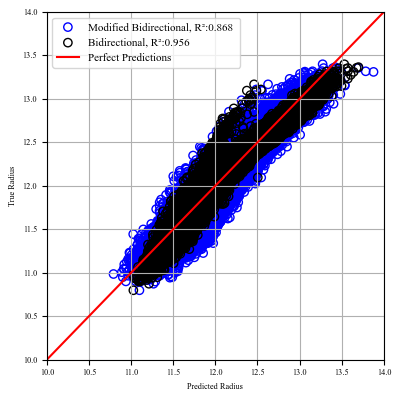}
\includegraphics[scale = .55]{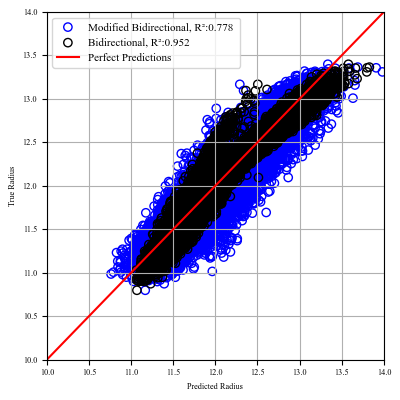}
\caption{Display of predicted radii from regression model against true radii from Bayesian inference. Dimensionless radius are transformed to have 
units of kilometers. Red line is the perfect fit line, i.e. if the prediction were exact against the true radii then all points would be on the red line. Blue circles indicate the predicted and true values for a given prediction of the model selected by the modified bidirectional feature selection process. Black circles are the predicted values for the selected model from normal bidirectional feature selection. Each plot shows the fits when different complexity terms are considered in the feature generation. Left shows fits of the fifth model when no cross terms are generated. Middle shows fits for the tenth model when cross terms up to degree three polynomial terms are generated . Right shows fits for the tenth model cross terms up to degree six are generated.   }
\label{fig:compFit}
\end{figure*}

In Fig.\ \ref{fig:deg6Dynam} we show the models selected when all cross terms up to degree six are generated and filtered by excessive correlation. The models are similar up to $M3^{6}$ where  
features $K_{sym}J_{0}^{2}$ and $LK_{0}^{4}$ are chosen first. 
However, here we note that the $R^{2}$ value for the model of which $R_{1.4}$  is a linear combination of  $K_{sym}J_{0}^{2}$  and $  LK_{0}^{4}$ is lower than the $R^{2}$ for the model with $R_{1.4}$  as a linear combination of only  $K_{sym}$ and $ L$ (which were the features selected by the selection algorithms used on  lower complexity data sets).  
At $M3^{6}$ the models found by the different step wise selection procedures begin to vary in $R^{2}$ scores and type of parameters selected. This difference is seen between selection algorithms and between the same selection algorithm used on different complexity data sets. 
For example, the modified process includes $J_{sym}$ at $M3_{mod}^{6}$, a deviance from the previous selecting algorithms used on features in a lower complexity feature set which did not  include $J_{sym}$ unless it was paired with another feature or as a reciprocal of itself. Moreover, after $M3_{mod}^{6}$ the modified algorithm chose variables which tended to have higher degree terms, increasing the complexity of the model, whereas the unmodified algorithm tend to choose the original EOS parameters first and then include some higher complexity terms. When the number of higher complexity terms increase the modified feature selection algorithm  produced a lower $R^{2}$ than when less high complexity terms are involved. The $R^{2}$ score when the degree six feature set was used is smaller than that when the degree two feature set is used. The unmodified algorithm's $R^{2}$ score was unaffected by a change in complexity of the feature set, only a change in the variables found.

Fig.\ \ref{fig:compFit} compares the regression-model predicted radii with the true radii (the  observation or actualization defined earlier) from the Bayesian inference. 
Black circles represent the predicted vs. true radius values when the model is selected by the original bidirectional feature selection algorithm, and blue circles are the models selected by the modified feature selection procedure. The leftmost panel gives the model fits when no cross terms are considered, middle panel are the fits when degree three terms are considered, and the rightmost panel is when cross terms up to degree 6 are considered. As the complexity of the feature set increases   both methods find models which better predict radius under 11.5 km and over 13.5 km. In the unmodified procedure the model accuracy as measured by the $R^{2}$ scores is maintained around 0.95 while the modified procedure finds models of varying $R^{2}$ scores. When no cross terms are considered the model achieves an accuracy around 77\%, introducing third degree cross terms leads to an approximately 10\% increase in $R^{2}$, and introducing sixth degree cross terms yields a less accurate model.

\begin{figure*}
\includegraphics[scale =.55]{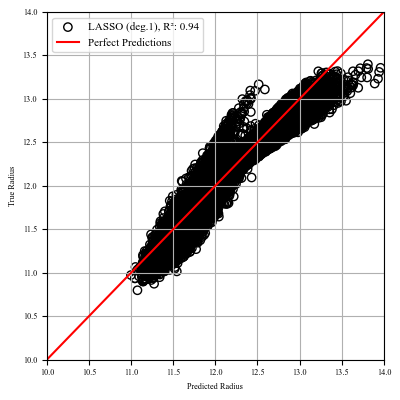}
\includegraphics[scale =.55]{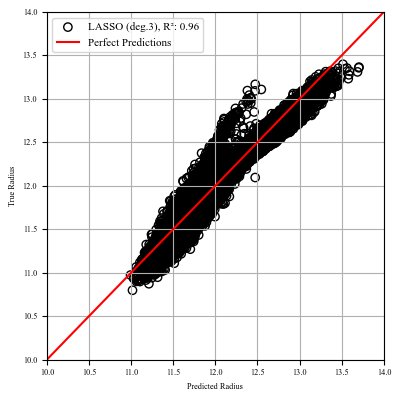}
\includegraphics[scale =.55]{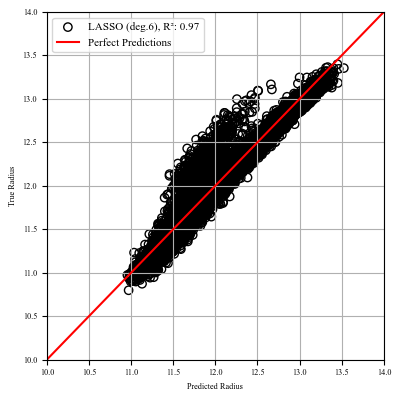}
\caption{ Predicted radii of LASSO regression selected models against the true radii from the Bayesian inference. Penalization is chosen to minimize AIC.  Left panel gives predictions when no cross terms are considered. Center panel gives the regression when at most degree three cross terms are considered. Right panel shows model when cross terms up to degree six are used.  }
\label{fig:lassComp}
\end{figure*}

The models that have more symmetric residuals are found by the modified bidirectional process. Models found by the regular algorithm are more accurate but fail to provide consistent information as one moves along the predicted radius line in red. The model which best combines accuracy and symmetric residuals  is shown in the middle panel when degree three cross terms are considered by the modified bidirectional algorithm. However, 
there is a slight bend in the predictions just under 12.5 km, which indicates some necessary non-linear terms may be missing from the selected model. This problem disappears when degree six terms are generated as shown in the rightmost plot.  


Imposing the condition (ii) in the feature selection process improves the distribution of the residuals around the prediction line and finds models which are sufficiently accurate. This corrects for problems in the models found by the regular forward feature selection. 
Namely, it reduces over predictions that occur at lower radii around 11.5 km and higher ones over 13.5 km and the inconsistent residuals when the radii are between the two extremes. The cost of imposing condition (ii) is model accuracy.  
Furthermore, the condition (ii) is less consistent as more cross terms are generated so care must be taken in determining the complexity of cross terms to use and what correlation to use for filtering. 

Regardless of the method and degree cross terms generated, $K_{sym}$ is the most heavily weighted parameter and the second most weighted one is $L$. This is strong evidence that $K_{sym}$ plays the most important role in predicting $R_{1.4}$ and that $L$ is slightly less important. Importantly, in none of the analysis does a cross term containing $L$ and $K_{sym}$ appear in the selected models. This is not certain proof their cross terms are not important, rather an indication that the cross terms do not provide any significantly new information when $K_{sym}$ and $L$ are already included. Notably, $J_{sym}$ is useful to increase the accuracy of the models selected. However, to generate a model with more symmetric residuals it appears only when many other parameters are already in the model or it is paired with another feature and scaled such that it produces more useful predictions. 

\subsection{Relative Importance of EOS Features According to LASSO Regression}
Fig. \ref{fig:lassComp} shows the models selected by LASSO regression with varying degrees of generated feature complexity. From left to right, the maximum generated degrees are one, three, and six, respectively. As the generated complexity increases the fits of LASSO regression improve qualitatively in the residuals and by the model selection statistics $R^{2}$ and AIC. 
Initially, with no cross terms considered  the distribution is similar to the bidirectional selection algorithms and there are clear bends and branches in the predictions which indicate the need for non-linear terms. 
In the middle plot where at most degree three cross terms are generated, there is a significant decrease in the bend and the branch shrinks closer to the main sequence of predictions. 
In the final model with generated degree six cross terms, the bends disappear. However, there is still a branch in the predictions when $R_{1.4}$ is around 12 km. Fortunately, the branch is linear and the true radii are closer to the prediction line. 

To obtain the fits shown in Fig.\ \ref{fig:lassComp}, the LASSO selection maintained a large number of features in its selection.  
For the case when degree three terms were generated, there were 17 features with coefficients greater than 0.0 and 16 features greater than 0.05. 
In the highest complexity feature set (right panel) there were 29 features with coefficients greater than 0.0 and 24 features with coefficients greater than 0.005.
The number of features from LASSO is nearly triple and double the number of parameters used by the bidirectional feature selection algorithms.

Table \ref{tab:lassParams} shows the coefficients for the features with the largest coefficients in the LASSO regression model which considered up to degree six cross terms. It is seen that $K_{sym}$ dominates the coefficients in the LASSO model when paired with other terms, especially when in cross terms involving $J_{0}^{2}$ and $J_{sym}^{2}$. If some feature vectors experience high multicollinearity the sizes of their coefficients in the regression model can vary greatly. Since  multicollinearity is evidence that features provide similar information. In this case, the effects of one feature's information can be spread to the coefficients of other features by a trade-off in coefficient size. Thus, it is possible that because $K_{sym}$ is included in many cross terms then many of the terms with $K_{sym}$ would experience exchanges in coefficient sizes and  decrease the coefficient of $K_{sym}$. Table \ref{tab:lassParams} demonstrates that this isn't the case, here the coefficient leading the lone $K_{sym}$ term is roughly the same proportion when no extra cross terms are used, as in Fig. \ref{fig:parModel}. Thus, cross terms with $K_{sym}$ provide information which is significant to $R_{1.4}$ apart from $K_{sym}$ as a lone term. 
Moreover, not explicit in Table \ref{tab:lassParams} the features of LASSO regression with non-zero coefficients were mostly composed of terms crossed with $K_{sym}$ and $L$. 
From the above analysis of LASSO regression, we would conclude that $K_{sym}$ is the most important parameter in determining the $R_{1.4}$. 

In comparison with the previous feature selection algorithms, LASSO finds that $K_{sym}$ is significantly more important than $L$. LASSO may also increases accuracy of the predictions if one is willing to accept the loss of interpretability in having more features and branch of predictions deviating from true values in an interval around the most probable radius of 12 km.

\begin{table}
\caption{Coefficient leading the indicated feature when cross terms up to degree six are generated.}
\vspace{0.1cm}
\label{tab:lassParams}

\begin{tabular}{c|c}
\hline
Feature & Coefficient (deg 6.) \\ \hline
$K_{sym}$ & 1.27 \\ 
$L$ & 0.91 \\
$K_{sym} J_{sym}^{2}$ & -0.63\\
$J_{sym}$ & 0.51 \\
$K_{sym} J_{0}^{2}$ & 0.32\\
$K_{sym}^{4}$ & -0.26 \\
$K_{sym} L^{2} J_{0}^{2}$ & -0.24 \\
$L J_{0}$ & 0.21 \\
$K_{0}$ & 0.19 \\ 
$J_{sym}^{5}$  & -0.16\\
\hline
\end{tabular}
\end{table}

\subsection{Relative Importance of EOS Features According to Neural Network Regression}
\begin{figure}
\centering
\vspace{-3.8cm}
\hspace{-2.7cm}
 \resizebox{0.6\textwidth}{!}{
\includegraphics[scale =1.]{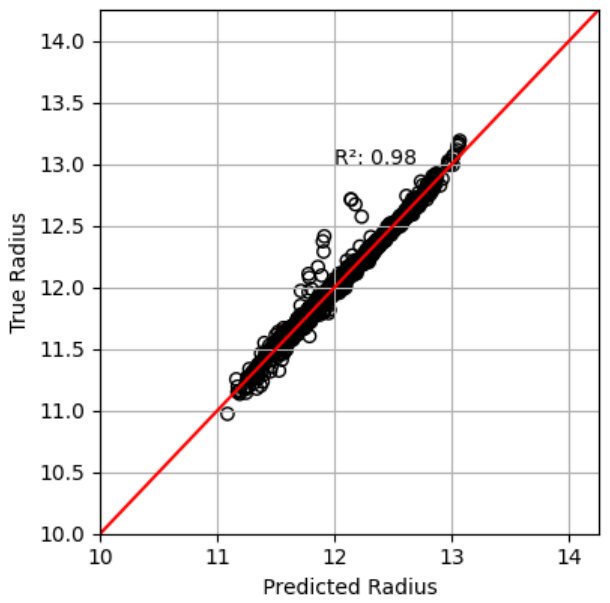}
}
\vspace{-4.1cm}
\caption{Neural network regression results on a testing subset of the original data.  Black circles are the predictions of the neural network against the true values. The red line indicates a perfect prediction for a calculated radius. Network was trained with sklearn package and a tanh activated hidden layer.  }
\label{fig:neuNet}
\end{figure}

Fig.\ \ref{fig:neuNet} shows neural network regression predictions against a test data set. 
The fit is nearly perfect with an $R^{2}$ score of 0.98 and no exceptional variations in the residuals except when around 12 km where some predictions are less than their true values. 
We also note slight under predictions when the true radius is around 13 km. Here the neural network regression technique with a single type of activation function and smaller number of layers provides the most accurate model for $R_{1.4}$. 

Shown in Table \ref{tab:featImp_nn} are the feature importance as measured by the change in MSE after permutation. We see that a permutation in $K_{sym}$ has the highest effect on MSE, indicating it is the most important parameter in the neural network model. After $K_{sym}$, a permutation in $L$ has the second highest effect in the model which is roughly 55\% the effect size  of permuting $K_{sym}$. The third most important feature is $J_{sym}$ with effect size 83\%  of the effect from permuting $L$ and 46\% the effect size from permuting $K_{sym}$. The permutation importance presents relatively strong evidence about the effects of $K_{sym}$ compared to $L$ and weaker evidence about the importance of $L$ when compared with $J_{sym}$ in modelling $R_{1.4}$. These findings help clarify the contradictory or ambiguous situation in the current literature which generally places large emphasis on the effects of $L$ in determining $R_{1.4}$ \cite{LRP2}. 

\begin{table}
\centering
\caption{Changes in model mean square error (MSE) from permutation. Higher MSE indicates a features which had larger weighting in the neural network, i.e. more important to the model. }
\vspace{0.2cm}
\label{tab:featImp_nn}
\begin{tabular}{c|c}
\hline Feature & Permutation MSE change \\\hline 
$K_{sym}$ & $2.88\pm 0.034
$\\
$L$  & $1.59 \pm .024$\\
$E_{sym}$& $0.011 \pm 0.00031$\\
$J_{sym}$ & $1.32 \pm 0.0204$ \\
$K_{0}$ & $0.131 \pm 0.0019$ \\
$J_{0}$  & $0.179 \pm 0.0026$\\ \hline
\end{tabular}
\end{table}

\begin{table*}
\raggedright
\caption{List of the best viable $R_{1.4}$ models as functions of the EOS parameters from the parametric regression-model building procedures (in terms of the reduced observable and variables). In parenthesis is the maximum degree cross term generated in the feature set used by the given method. The modified bidirectional feature selection models were included because they maintained more consistent residuals than the regular feature selection process. Only the LASSO model which considered all filtered degree six terms is listed with many smaller coefficients that can be disregarded. }
\vspace{0.2cm}
\label{tab:all_models}
\begin{tabular}{|c|c|m{20em}|c|c|}

\hline Selection Method (Deg. Considered.) & $\#$ of Variables& R$_{1.4}$ as function of reduced EOS parameters & AIC & $R^{2}$\\ \hline

Modified Bidirectional (1)  & 3 & $0.735K_{\rm sym} + 0.7199 L + 0.144 K_{0} $& -89186 & 0.727\\ \hline 

Modified Bidirectional (2)  & 5 & $ 0.896 K_{\rm sym} + 0.675 L + 0.556 (LJ_{\rm sym})^{\frac{1}{3}} + 0.232 K_{0} + 0.334 J_{0} $ & -123195 & 0.8343 \\ \hline 

Modified Bidirectional (3)  & 7 & $1.262 K_{\rm sym} + 0.696 L - 0.517 K_{\rm sym}J_{sym}^{2} + 0.34052 K_{\rm sym} J_{0}^{2} - 0.23537 (K_{\rm sym}E_{\rm sym}^{2})^{\frac{1}{3}} + 
0.218 L^{2}J_{\rm sym} + 0.153 J_{0}^{2}$ & -134567 & 0.859 \\\hline 

Modified Bidirectional (4)  & 10 & 
0.711$K_{\rm sym} J_{0}^{2}$    
+ 0.287$1/L K_{0}^3$    
+  0.619 $J_{\rm sym} $    
-0.384$(K_{s}J_{0})^{1/3}+$     
+0.941$L+$    
 0.517 $K_{0}$    
 -0.105 $K_{0}E_{\rm sym}+$    
-0.190 $L^2J_{\rm sym}$     
-0.0914 $1/L^4+$     
 0.065 $(L J_{\rm sym})^{1/3}$    & -102773  & 0.776 \\\hline

Modified Bidirectional (6)  & 10 & 
0.740 $K_{\rm sym}  J_{0}^{2}$  +0.643 $J_{\rm sym}$ 
 + 0.193 $L K_{0}^4$ +    0.567 L  
-0.324 ($K_{\rm sym}J_{0})^{\frac{1}{3}}$ -0.247 $ L^2 J_{\rm sym}$  0.199$K_{0}$ 
-0.082 $E_{\rm sym}$   
+0.087$L^6$    
+0.066$L^2 J_{\rm sym}^4$   
& -103119  & 0.778 \\ \hline 

LASSO (6) & 24 & 1.27 $K_{\rm sym}$ 
+0.91$L$ -0.630$K_{\rm sym} J_{\rm sym}^{2}$
+ 0.511$J_{\rm sym}$ +0.312$K_{\rm sym}  J_{0}^{2}$
 -0.259 $K_{\rm sym}^{4}$  
 -0.241 $K_{\rm sym}  L^2 J_{0}^2$ +0.219 $L J_{0}$  + 0.195$K_{0}$ -0.163$J_{\rm sym}^{5}$  -0.098$(K_{\rm sym}  E_{\rm sym}^{2})^{\frac{1}{3}}$ -0.085$L^2 J_{\rm sym}^4 $
+ 0.076$L K_{0}^4$ + 0.073 $J_{\rm sym}^2 J_{0}^{2}$ 
-0.065$E_{\rm sym}$ -0.051$L J_{0}^2 E_{\rm sym}^2$ -0.050$L^{2} J_{\rm sym}$  
 -0.036$L E_{\rm sym}^{3}$  +0.036$(K_{\rm sym} J_{0})^{\frac{1}{3}}$ -0.035$K_{\rm sym}  E_{\rm sym}^5$ 
 -0.032 $(J_{\rm sym} K_{0}^2 J_{0})^{\frac{1}{3}}$  
+ $(L J_{0} )^{\frac{1}{3}} E_{\rm sym}$ 
-0.026$K_{\rm sym} L^{2}$ + 0.016$L^6$  -0.006$1/L^4$ 
 +0.004 $J_{0}E_{\rm sym}^{5}$& -243405 & 0.971 \\ \hline
\end{tabular}
\end{table*}

\subsection{Identifying the Most Influential EOS Features and Selecting the Most Accurate $R_{1.4}$ Models}
To this end, it is useful to summarize here the key advantages and disadvantages of each technique in developing the regression models for $R_{1.4}$. What do they predict in common about the most influential EOS features and the most accurate $R_{1.4}$ models?

To our best knowledge, there is no proper prior choice of bidirectional selection techniques, i.e. modified or unmodified for a given data set. The unmodified bidirectional selection techniques provide quicker convergence to accurate models and can incorporate a lower number of parameters that maintain high model accuracy. However, the fits show that the regular procedure have large changes in residual variation and consistency. Given a prediction for $R_{1.4}$ below 11.5 km or over 13 km, a modeller using the unmodified bidirectional selection technique would be confident that their results are significantly over predicting the radius, but would be assured that between these values their predictions are more accurate. The modified selection algorithm generally solves the problems of the unmodified one, creating more consistent residuals around the prediction. In this case, it would be impossible for a modeller to be confident they are over predicting unless they are strictly at the tail of the prediction interval. However, there is a trade-off in the accuracy of the models and consistency of variable selection as complexity increases. Thus, we should note that employing either of these techniques extra caution should be taken in analyzing and improving the residuals. 

Between the bidirectional selection and LASSO regression, LASSO provides an efficient method to determine which features are most important, considering all of them simultaneously. 
In the analysis, this generally led LASSO regression to maintain a larger subset of features than either of the bidirectional selection techniques. 
Since, the fits of LASSO regressions improve the distribution of the residuals when higher complexity feature sets were utilized then we may assume both bidirectional feature techniques terminate too early and there are likely few missing features in their models. However, due to the iterative construction it is possible that the parameters in bidirectional selection are correlated with the extra parameters LASSO maintains. Importantly, since the fits from LASSO regression improve with higher complexity terms which include $K_{sym}$ then there are likely many important cross terms between $K_{sym}$ and other EOS parameters that present vital information not contained in $K_{sym}$ alone. The clear trade-off with choosing LASSO regression is a loss in interpretability from a significant increase in the number of parameters. 

Of all the modeling techniques, neural network regression provides the best fit and generalizability, even with a relatively small and simple network. However, there is no way to get a parametric model of $R_{1.4}$. Thus, to provide an intuitive model of $R_{1.4}$ it is suggested that one of the parametric techniques (LASSO or bidirectional selection) discussed earlier be used instead of a neural network. However, if the accuracy is most important then a neural network should be used.

We demonstrated the effectiveness of each regression-model building strategy when applied to a posterior EOS data set that has dependencies induced by the Bayesian inference. Combining the analyses we can determine the most important features in predicting $R_{1.4}$. The bidirectional feature selections provide relatively weak evidence for the importance of $K_{sym}$ compared with $L$. The coefficients leading $K_{sym}$ and $L$ are stronger evidence that $K_{sym}$ is more important, especially if one considers the cross terms involving $K_{sym}$ and $L$. The neural network, which has the highest accuracy provides the strongest evidence for the importance of $K_{sym}$ in comparison to $L$, where the difference in effect size among the MSE statistics are much greater than when comparing the coefficients of $K_{sym}$ and $L$. Although these numbers cannot be compared directly they all provide significant evidence in favor of $K_{sym}$ being the most important parameter in determining $R_{1.4}$.

Notably, $J_{sym}$ was found to also play a major role in predicting $R_{1.4}$. This information is evident in all three feature selection techniques. The parametric models which did not include $J_{sym}$ tended to perform worse and the neural network regression depended on $J_{sym}$ nearly as much as it did on $L$. We suggest that any model for $R_{1.4}$ should then include $J_{sym}$ and that many viable models can be constructed from combinations of $K_{sym}$, $L$ and $J_{sym}$ (i.e., the density dependence of nuclear symmetry energy) alone. This is a different conclusion than would be made if analysis was only made on linear correlations between the EOS parameters and $R_{1.4}$, since $J_{0}$ has greater linear correlation with $R_{1.4}$ than $J_{sym}$ but $J_{sym}$ is significantly more important to predict $R_{1.4}$ regardless of the regression modelling technique used. 

We emphasize that the above observations are not due to the different variances of the EOS parameters since (i) we normalized the standard deviation to unity in preprocessing the EOS features and  (ii) the variance of $L$ is significantly smaller than the variance of $J_{0}$ and was found to have a much larger effect.  If only the variances dictated model selections, then it would be expected that $J_{0}$ have a larger effect on $R_{1.4}$ than $L$.

Table \ref{tab:all_models} summarizes the above discussion, showing the best models found from each procedure and showing their $R^{2}$ and AIC scores. The best models included were from applying the modified bidirectional selection techniques since they provide the most consistent distribution of residuals, which we take to be more important than the raw $R^{2}$ score. Moreover, of the bidirectional feature selections the best model was obtained when the maximum degree features generated were degree three. They had the highest $R^{2}$ and AIC scores and maintained more consistent residuals. The best overall model by AIC, residuals, and $R^{2}$ was found by the LASSO regression when the highest degree features in the generated list were degree six, although only a few degree six terms were included in the model, and none with high coefficients leading them.

\begin{table}
\raggedright
\caption{The $R_{1.4}$ values in unit of Km predicted by using the 9 unified RMF EOSs with the listed $K_0$, L and $K_{\rm sym}$ parameters in unit of MeV from Ref. \cite{Fortin} in comparison with the predicted $R_{1.4}$-RL3 values.} 
\vspace{0.2cm}
\label{R14-comp}
\begin{tabular}{|c|c|c|c|c|c|}
\hline EOS &$K_0$ & L &$K_{\rm sym}$ & $R_{1.4}$ & $R_{1.4}$-RL3\\ \hline
NL3 &271.6 & 118.9 & 101.6& $14.63\pm 0.71$&14.45 \\ \hline 
NL3$\omega\rho$&271.6&55.5&-7.6&$13.75\pm 0.10$&12.80\\ \hline 
GM1&300.7&94.4&18.1&$13.76\pm 0.42$&13.81\\ \hline 
DDME2&250.9&51.2&-87.1&$13.27$&12.34\\ \hline 
TM1&281.2&111.2&33.8&$14.37$&13.87\\ \hline 
DDH$\delta$&240.3&48.6&91.4&$12.5$&12.85\\ \hline 
DD2&242.6&55&-93.2&$13.1$&12.36\\ \hline 
BSR2&239.9&62.0&-3.1&$13.4$&12.80\\ \hline 
BSR6&235.8&85.7&-49.6&$13.7$&13.10\\ \hline 
\end{tabular}
\end{table}
\begin{figure}
\centering
\includegraphics[scale = .35]{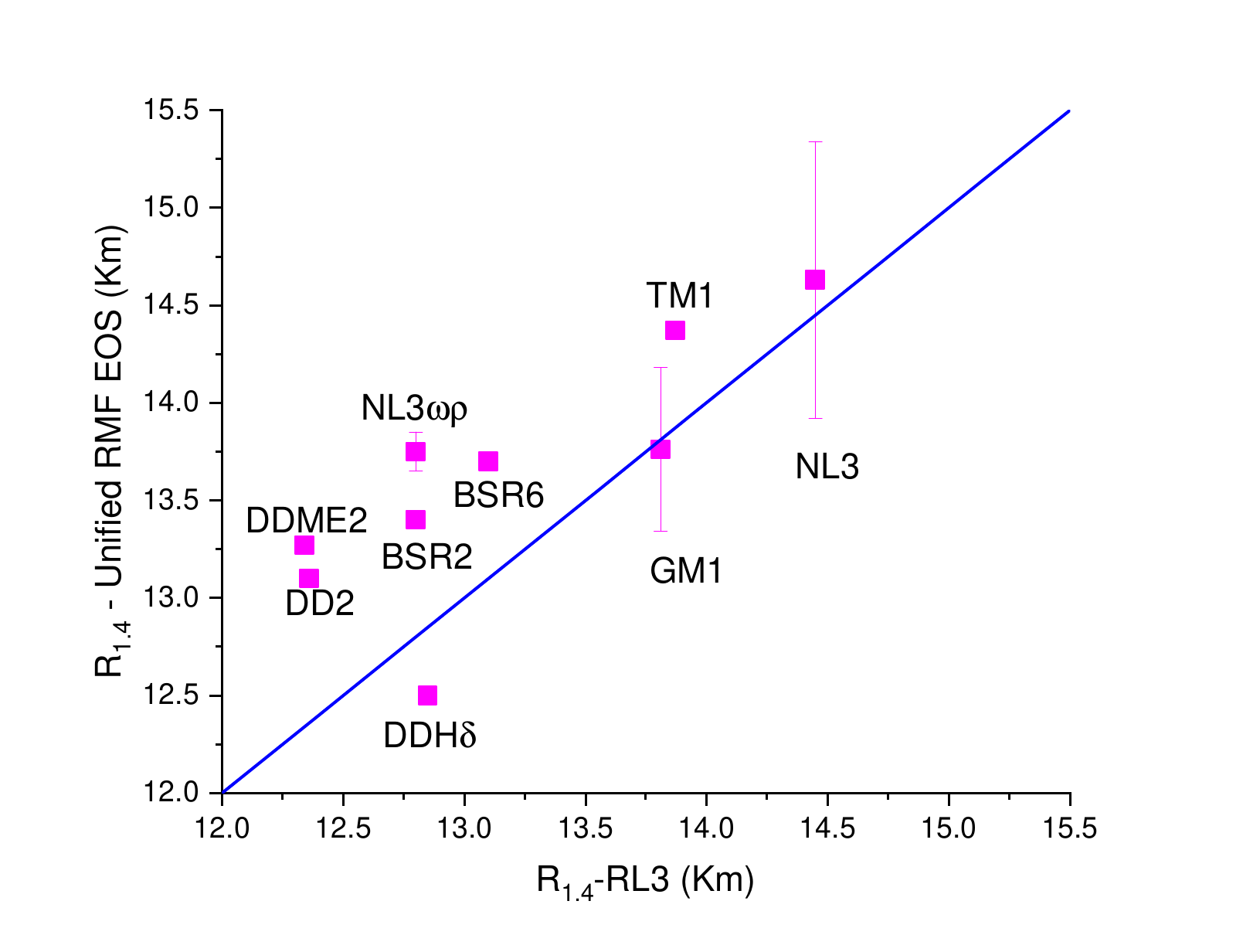}
\caption{A scatter plot of the $R_{1.4}$ values (vertical axis) predicted by using the 9 unified RMF EOSs from Ref. \cite{Fortin} versus the predicted $R_{1.4}$-RL3 values (horizntal axis) by using the Eq. (\ref{eqn:r14_rl3}). Any points lying along the blue line are a perfect match between the two values.}
\label{fig:r14_comp}
\end{figure}
\subsection{Comparisons with predictions of 9 unified RMF EOS models}
To this end, we emphasize again that our work is based on analyzing the large ensemble of posterior EOS models and their predictions for $R_{1.4}$ from earlier Bayesian analyses of astrophysical data \cite{Xie19,Xie20a,xi_Li_2021}. Needless to say, we did not consider all uncertainties known and the neutron star model we used is the minimum model. To our best knowledge, there is no consensus in the community about what constitute realistic models of neutron stars as there are just too many unknowns about their internal structures and the observational data is very limited. By definition, all EOS models 
generated through the Bayesian analyses have certain probabilities to reproduce all the observational data considered. In this sense, our empirical formulas for $R_{1.4}$ extracted from applying the modern regression techniques to these Bayesian posterior EOS models and their predictions are realistic in terms of reproducing the observational data. How good are our empirical formulas 
when compared to predictions of more microscopic EOS models that go beyond the minimum model for neutron stars in various degrees? The answers to this question are 
model dependent as there are  many such models in the literature. With some biases we choose to compare our simplest 3-parameter formula listed in \ref{tab:all_models} with predictions of the nine {\it unified} Relativistic Mean Field (RMF) model predictions by Fortin {\ et al.} in Ref. \cite{Fortin}. These {\it unified} EOSs have all segments (outer crust, inner crust, liquid core) calculated starting from the same nuclear interaction. Moreover, for three of the nine RMF EOSs, effects of using 5 or 6 different ways to match the crust and core are considered.

The results shown in table \ref{tab:all_models} can be used to quickly generate a set of $R_{1.4}$ values once the EOS parameters are known by transforming the features to their original scale according to Eq. (\ref{redu}) and Table \ref{tb:eosStats}. For example, the three-parameter expression for the scaled radius $R_{1.4} = 0.735K_{\rm sym} + 0.7199 L + 0.144 K_{0}$ in terms of the scaled $K_{sym}$, $L$, and $K_{0}$ in Table \ref{tab:all_models} can be re-written all using the original variables as
\begin{eqnarray}
R_{1.4} &=& 12.10+0.38[0.735(K_{\rm sym}+158.97)/80.45\nonumber\\ 
&+& 0.7199(L-54.13)/13.63\nonumber\\ 
&+& 0.144(K_{0}-239.95)/11.52].\label{eqn:r14_rl3}
\end{eqnarray}
We denote this radius as $R_{1.4}$-RL3 in the following discussions to distinguish it from those one may get using other expressions involving more EOS parameters as listed in Table \ref{tab:all_models}. While the $R_{1.4}$-RL3 gives a relatively small coefficient of determination $R^2=0.727$, it is the simplest and currently the most useful one because most of the works in the literature have only published the $K_{sym}$, $L$ and $K_{0}$ values without giving any information bout the higher-order EOS parameters. 

Table \ref{R14-comp} lists the relevant RMF EOS model parameters as well as their corresponding $R_{1.4}$ (from RMF EOS) and $R_{1.4}$-RL3 values (from our empirical formula of Eq. (\ref{eqn:r14_rl3})). For the RMF EOSs in the first three rows, effects of using different ways to match the crust-core EOSs were studied in Ref. \cite{Fortin}. The maximum effects of these matching methods are indicated by the error bars of the predicted $R_{1.4}$ values. Fig. \ref{fig:r14_comp} further compares the $R_{1.4}$ and $R_{1.4}$-RL3 values. We notice that any points lying along the blue line reflect a perfect match between the two values. We see the our empirical radius formula with three variables is consistent in reconstructing the predictions from the RMF EOSs and is especially consistent at higher radii. We note particularly good agreement with $R_{1.4}$ from the $GM1$ and $NL3$ EOS. 

Despite of the fact that some of the EOS models give $K_{\rm sym}$ and L values that are outside the currently known 68\% confidence boundaries of these two parameters based on many Bayesian analyses of astrophysical data ( see, e.g., the Table \ref{tb:eosStats} and the systematics of $L=57.7\pm 19$ MeV and $K_{\rm sym}=-107\pm 88$ MeV based on 24 independent Bayesian analyses of LIGO/VIRGO data \cite{LIBA21}), the $R_{1.4}$-RL3 seems to at least capture the main features of the RMF model predictions. The physics underlying some of these 9 RMF models is quite different in several aspects. They also have very different assumptions at their conception, leading to the rather different EOS parameters as illustrated in Table \ref{R14-comp}. The agreement between the $R_{1.4}$ and $R_{1.4}$-RL3 values is very reasonable (differ by less than 7\% in comparison with the current accuracy of more than 10\% in the radius measurement of neutron stars). This is suggestive that it is the proper combination of the EOS parameters (rather than their individual values) that is most important for determining the radius of a canonical neutron star. Thus, the above comparison supports the conclusion that our empirical radius formula for canonical neutron stars has the right combination of nuclear EOS parameters.   

\section{Conclusions}\label{s5}
To identify the most important EOS features determining the radius $R_{1.4}$ of canonical NSs, we applied the three state-of-the-art regression-model building methodologies: bidirectional step-wise feature selection, LASSO regression, and neural network regression to posterior EOSs inferred from Bayesian analyses of NS observational data. 

We extracted the best $R_{1.4}$ models with varying accuracy and complexity in terms of the EOS features. Given new constraints or realizations of the EOS parameters from future experiments, one can immediately see their impact on $R_{1.4}$ using these empirical formulas without having to construct new EOS models for NSs and solving again the TOV equations. All three regression methods agree about the most important predictors of $R_{1.4}$. In order of decreasing importance, these are $K_{sym}$, $L$, $J_{sym}$, $J_{0}$, $K_{0}$ and $E_{sym}(\rho_0)$. \\

\section*{Acknowledgments} We would like to thank Profs. Wen-Jie Xie and Nai-Bo Zhang for helpful discussions. 
This work was supported in part by the U.S. Department of Energy, Office of Science, under Award Number DE-SC0013702.



\begin{thebibliography}{99}
\bibitem{LRP1}
A.~Lovato, \textit{et al.}
[arXiv:2211.02224 [nucl-th]].

\bibitem{LRP2}
A.~Sorensen, \textit{et al.}
[arXiv:2301.13253 [nucl-th]].

\bibitem{Baiotti}L. Baiotti, Prog. Part. Nucl. Phys., {\bf 109}, 103714 (2019).

\bibitem{BALI19}B.A. Li, P.G. Krastev, D.H. Wen, and N.B. Zhang, Euro. Phys. J. A, {\bf 55}, 117 (2019).

\bibitem{Rai19}
C.~A.~Raithel,
Eur. Phys. J. A \textbf{55} no.5, 80 (2019).

\bibitem{Capano20} C. D. Capano, I. Tews, S. M. Brown, B. Margalit, S. De, S. Kumar, D. A. Brown, B. Krishnan, S. Reddy, Nature Astronomy, {\bf 4}, 625 (2020).

\bibitem{Kat20} Katerina Chatziioannou, General Relativity and Gravitation, {\bf 52}, 109 (2020).

\bibitem{AngLi}  A. Li, Z. -Y. Zhu, E. -P. Zhou, J. -M. Dong, J. -N. Hu, C. -J. Xia, Journal of High Energy Astrophysics, {\bf 28}, 19 (2020).

\bibitem{Huth}
S.~Huth, P.~T.~H.~Pang, I.~Tews, T.~Dietrich, A.~L.~F\`evre, A.~Schwenk, W.~Trautmann, K.~Agarwal, M.~Bulla and M.~W.~Coughlin, \textit{et al.}
Nature \textbf{606}, 276 (2022).

\bibitem{Bom91} I. Bombaci, U. Lombardo, Phys. Rev. C \textbf{44}, 1892 (1991).

\bibitem{Sot2}
H.~Sotani and H.~Togashi,
Phys. Rev. D \textbf{105}, 063010 (2022).

\bibitem{Sot22}
H.~Sotani and S.~Ota,
Phys. Rev. D \textbf{106}, 103005 (2022).

\bibitem{Sot3}
H.~Sotani and T.~Naito,
Phys. Rev. C \textbf{107}, 035802 (2023).

\bibitem{JXu2} J. Xu, L.W. Chen, B.A. Li, H.R. Ma, Astrophys. J. \textbf{697}, 1549 (2009).

\bibitem{Xu09}J.~Xu, L.W.~Chen, B.A.~Li, H.~R.~Ma,
Phys. Rev. C \textbf{79}, 035802 (2009).

\bibitem{New12} W.G. Newton, M. Gearheart, B.A. Li, Astrophys. J. Supplement Series \textbf{204}, 9 (2013).

\bibitem{New14}
W.G.~Newton, J.~Hooker, M.~Gearheart, K.~Murphy, D.~H.~Wen, F.~J.~Fattoyev and B.~A.~Li,
Eur. Phys. J. A \textbf{50}, 41 (2014).

\bibitem{Pro14} C. Provid\^{e}ncia, S.S. Avancini, R. Cavagnoli, S. Chiacchiera, C. Ducoin, F. Grill, J. Margueron, D.P. Menezes, A. Rabhi, I. Vida\~{n}a, Euro. Phys. J. A \textbf{50}, 44 (2014).

\bibitem{Tews17} I. Tews, J.M. Lattimer, A. Ohnishi, E.E. Kolomeitsev, Astrophys. J 848, 105 (2017).

\bibitem{India17} C. Mondal, B.K. Agrawal, J.N. De, S.K. Samaddar, M. Centelles, X. Vi\~{n}as, Phys. Rev. C \textbf{96}, 021302(R) (2017).

\bibitem{Holt}J. W. Holt and Y. Lim, Physics Letters B {\bf 784}, 77 (2018).

\bibitem{Fra-Crust2} T. Carreau, F. Gulminelli, J. Margueron, Eur. Phys. J. A {\bf 55}, 188 (2019).

\bibitem{Magn}B.A.~Li and M.~Magno,
Phys. Rev. C \textbf{102}, 045807 (2020).

\bibitem{Zhang2018} N.B. Zhang, B.A. Li, J. Xu, Astrophys. J. \textbf{859}, 90 (2018).

\bibitem{NBZ-JPG}
N.B.~Zhang and B.A.~Li,
J. Phys. G \textbf{46}, 014002 (2019).

\bibitem{Xie19}W.J. Xie and B.A. Li,  Astrophys. J. 883, 174 (2019).
\bibitem{Xie20a}W.J. Xie and B.A. Li, Astrophys. J. 899, 4  (2020).
\bibitem{xi_Li_2021} W.J. Xie and B.A. Li, 
Phys. Rev. C 103, 035802 (2021).

\bibitem{LIGO18} B.P. Abbott et al., Phys. Rev. Lett. {\bf 121}, 161101 (2018).

\bibitem{De18} S. De, D. Finstad, J.M. Lattimer, D.A. Brown, E. Berger, and C.M. Biwer,  Phys. Rev. Lett., {\bf 121}, 091102 (2018).

\bibitem{Lattimer14} J.M. Lattimer and A. W. Steiner,  Eur. Phys. J. A, 50, 40 (2014).

\bibitem{Miller19}M.C. Miller et al., ApJL, 887, L24 (2019).

\bibitem{Riley19}T.E. Riley et al., ApJL, 887, L21 (2019).

\bibitem{Negele73} J.W. Negele and D. Vautherin, Nucl. Phys. A {\bf  207}, 298 (1973).
\bibitem{Baym1971} G. Baym, C.J. Pethick, P. Sutherland, Astrophys. J. \textbf{170}, 299 (1971).
\bibitem{Lattimer00} J.M. Lattimer, M. Prakash, Phys. Rep. \textbf{333}, 121 (2000).

\bibitem{Kubis} S. Kubis, Phys. Rev. C \textbf{76}, 025801 (2007).

\bibitem{Zhang19} N.B. Zhang and B.A. Li, Euro. Phys. J. A \textbf{55}, 39 (2019).
\bibitem{Zhang19b} N.B. Zhang and B.A. Li, Astrophys. J. \textbf{879}, 99 (2019).
\bibitem{Zhang20} N.B. Zhang and B.A. Li, Astrophys. J. \textbf{902}, 38 (2020).
\bibitem{Zhang20b} N.B. Zhang and B.A. Li, Astrophys. J. \textbf{893}, 61 (2020).
\bibitem{Zhang21} N.B. Zhang and B.A. Li, Astrophys. J. \textbf{921}, 111 (2021).
\bibitem{Zhang23} N.B. Zhang and B.A. Li,
Phys. Rev. C \textbf{108}, 025803 (2023).
\bibitem{Oppenheimer39} J. Oppenheimer and G. Volkoff, Phys. Rev. \textbf{55}, 374 (1939).

\bibitem{Mond23}
C.~Mondal and F.~Gulminelli,
Phys. Rev. C \textbf{107}, 015801 (2023).


\bibitem{LIBA21} B.A. Li, B.J. Cai, W.J. Xie, N.B. Zhang, Universe \textbf{7}, 182 (2021).

\bibitem{NBZ23}N.~B.~Zhang and B.~A.~Li,
Eur. Phys. J. A \textbf{59}, 86 (2023).

\bibitem{eft}C. Drischler, R.J.Furnstahl, J.A.Melendez, D.R.Phillips, Phys. Rev. Lett. {\bf 125},202702 (2020).

\bibitem{verma} J. P. Verma and A.-S. G. Abdel-Salam, Testing Statistical Assumptions in Research (Wiley, Hoboken, NJ, 2019).

\bibitem{neter_regression} J. Neter, W. Wasserman, and M. H. Kutner, Applied Linear Regression Models, 2nd ed (Irwin, Homewood, Ill, 1989).

\bibitem{casella_stats}G. Casella and R. L. Berger, Statistical Inference, 2nd ed (Thomson Learnie, CA, 2002), Australia; Pacific Grov.

\bibitem{buhl_lasso} P. Bühlmann and S. van de Geer, Statistics for High-Dimensional Data: Methods, Theory and Applications (Springer, Berlin Heidelberg, 2011).

\bibitem{akaike_aic} H. Akaike, Information Theory and an Extension of the Maximum Likelihood Principle, in Breakthroughs in Statistics. 1: Foundations and Basic Theory (Springer, New York Berlin Heidelberg, 1992), pp. 610–624.

\bibitem{murphy_machine} K. P. Murphy, Machine Learning: A Probabilistic Perspective (MIT Press, Cambridge, MA, 2012).


\bibitem{scikit} F. Pedregosa et al., Scikit-Learn: Machine Learning in Python, (2012).

\bibitem{wackerly_stats}D. D. Wackerly, W. Mendenhall, and R. L. Scheaffer, Mathematical Statistics with Applications, 6th ed (Duxbury, Pacific Grove, CA, 2002).

\bibitem{Fortin}
M.~Fortin, C.~Provid\^{e}ncia, A.~R.~Raduta, F.~Gulminelli, J.~L.~Zdunik, P.~Haensel and M.~Bejger,
Phys. Rev. C \textbf{94}, 035804 (2016).

\end{thebibliography}
\end{document}